\documentclass{svjour3}

\begin{document}

\title{Galactic tide}
\author{J. Kla\v{c}ka}
\institute{Faculty of Mathematics,
Physics and Informatics, Comenius University \\
Mlynsk\'{a} dolina, 842 48 Bratislava, Slovak Republic \\
\email{klacka@fmph.uniba.sk}}

\date{}

\authorrunning{J. Kla\v{c}ka}
\titlerunning{Galactic tide}
\maketitle

\begin{abstract}
Equation of motion for the galactic tide is derived under the assumption
of cylindrically symmetric gravitational potential of the Galaxy.
The paper considers galactic tide both for the galactic plane
$x-$ and $y-$ components and also for the normal $z-$ component.
The $x-$ and $y-$ components of the acceleration come not only
from the $x-$ and $y-$ components of the position of a body, but also from
its $z-$component of the position vector.

Values of the Oort constants are
$A$ $=$ (14.2 $\pm$ 0.5) ~ $km ~s^{-1} ~kpc^{-1}$ and
$B$ $=$ ($-$ 12.4 $\pm$ 0.5) ~ $km ~s^{-1} ~kpc^{-1}$.
Mass density in the solar neighborhood, 30 $pc$ above the galactic
equatorial plane, equals to (0.117 $\pm$ 0.005) $M_{\odot}~ pc^{-3}$.
The result for the acceleration is written in the
form easily applicable to Solar System studies, to
the evolution of comets in the Oort cloud.

\keywords{Galaxy \and Oort constants \and Oort cloud \and Equation of motion}

\end{abstract}

\section{Introduction}
Global galactic gravitational field influences relative motion of two close
bodies in the form of galactic tide. This relative motion is important in
Solar System studies on evolution of comets within the Oort cloud of comets.
This paper derives the equation of motion for galactic tide under the
assumption of cylindrically symmetric gravitational potential of the Galaxy.
The effect of the normal $z-$component is well understood in the case
when motion of one of the bodies is fixed to galactic equatorial plane
(see, e.g., Mihalas and McRae Routly 1968, pp. 221-222; analytic approach
to the secular evolution of orbital elements in the gravitational field
of the Sun is presented in Kla\v{c}ka and Gajdo\v{s}\'{\i}k 1999).
Acceleration in the galactic plane comes not only from the $x-$ and $y-$
components of the galactic field, but also from the $z-$component of the
galactic components. Similarly, acceleration in the $z-$component comes also
from the $x-$ and $y-$ components of the galactic field acceleration/tide.
The acceleration in the plane parallel to the galactic equatorial plane
and the acceleration normal to the galactic equatorial plane
are presented in the relevant general form for two sets of galactic
components: one case corresponds to Dauphole et al. (1996) potential model
for galactic bulge, disk and halo, the other case considers potential galactic
bulge of Dauphole et al. (1996), galactic disk is represented by
mass density function (Maoz 2007) and galactic halo is given by flat rotation
curve.

\section{Values of the Oort constants}
In order to have a realistic model for galactic gravitational force in disposal,
one needs to know also the values of Oort constants. They refer for the
region of the Sun, in galactocentric distance about 8 kpc.

There is relatively great uncertainty in the values of the Oort constants.
The values with errors are given in Table 1, where also sources of the
values are given.

\begin{table}[h]
\centering
\begin{tabular}{|c|c|c|}
\hline
$A$ & $B$ & $source$  \\
\hline
[$km ~s^{-1} ~kpc^{-1}$] & [$km ~s^{-1} ~kpc^{-1}$] &  \\
\hline
\hline
15.0 $\pm$ 0.8 & $-$ 10.0 $\pm$ 1.2 & Kulikovskij (1985, p. 96) \\
\hline
14.4 $\pm$ 1.2 & $-$ 12.0 $\pm$ 2.8 & Kerr and Lynden-Bell (1986) \\
\hline
11.3 $\pm$ 1.1 & $-$ 13.9 $\pm$ 0.9 & Hanson (1987) \\
\hline
14.8 $\pm$ 0.8 & $-$ 12.4 $\pm$ 0.6 & Feast and Whitelock (1997) \\
\hline
14.5 $\pm$ 1.5 & $-$ 12 $\pm$ 3 & Barbieri (2007, p. 167) \\
\hline
\end{tabular}
\caption{Values of the Oort constants taken from various sources.}
\label{tab:1}
\end{table}

The values presented by Hanson (1987) are given also by
Olling and Merrifield (1998), the values presented by
Feast and Whitelock (1997) can be found also in
Sparke and Gallagher (2007, pp. 92-93) and
Carroll and Ostlie (2007, p. 913).

The scatter of the values of $A$ and $B$ is large.
The best estimates of the quantities $A$ and $B$ are the weighted averages
\begin{eqnarray}\label{1}
A_{wav} &=& \frac{\sum_{i=1}^{N} w_{A~i}~ A_{i}}{\sum_{i=1}^{N} w_{A~i}} ~,
\nonumber \\
B_{wav} &=& \frac{\sum_{i=1}^{N} w_{B~i} ~B_{i}}{\sum_{i=1}^{N} w_{B~i}} ~,
\end{eqnarray}
where the weight $w_{A~i}$ ($w_{B~i}$) of each measurement is the reciprocal
square of the corresponding uncertainty
\begin{eqnarray}\label{2}
w_{A~i} &=& 1 / \sigma_{A~i}^{2} ~,
\nonumber \\
w_{B~i} &=& 1 / \sigma_{B~i}^{2} ~, ~~~i = 1, 2, ..., N
\end{eqnarray}
(Taylor 1997, pp. 175-176) and the uncertainties in $A_{wav}$ and $B_{wav}$
are
\begin{eqnarray}\label{3}
\sigma_{A_{wav}} &=& 1 / \sqrt{\sum_{i=1}^{N} w_{A~i}} ~,
\nonumber \\
\sigma_{B_{wav}} &=& 1 / \sqrt{\sum_{i=1}^{N} w_{B~i}} ~.
\end{eqnarray}
Using the above presented values, we can calculate, on the basis of
Eqs. (1)-(3), the most probable values of $A$ and $B$ (we omit the index
``wav''):
\begin{eqnarray}\label{4}
A &=& ( 14.2 \pm 0.5 ) ~ km ~s^{-1} ~kpc^{-1} ~,
\nonumber \\
B &=& ( -~ 12.4 \pm 0.5 ) ~ km ~s^{-1} ~kpc^{-1} ~.
\end{eqnarray}
The values are consistent with the statements that ``$A$ has a value of about
14 $km ~s^{-1} ~kpc^{-1}$'' and ``$B$ takes a value around
$-$ 12 $km ~s^{-1} ~kpc^{-1}$'' (Phillipps 2005, p. 101). However,
these values significantly differ from the IAU recommended values
$A$ $=$ 15 ~ $km ~s^{-1} ~kpc^{-1}$,
$B$ $=$ $-$ 10 ~ $km ~s^{-1} ~kpc^{-1}$
(Karttunen et al. 2007, p. 358). Also the results obtained from
the model of Dauphole et al. (1996) are not fully consistent
with the results presented in Eq. (4).
The model of Dauphole et al. (1996) yields
$A$ $=$ 14.2 ~ $km ~s^{-1} ~kpc^{-1}$,
$B$ $=$ $-$ 13.9 ~ $km ~s^{-1} ~kpc^{-1}$.
Similarly, the cepheids data of Karimova and Pavlovskaja (1974)
(see also Kulikovskij 1985, p. 98) yield rotation curve
$v(R)$ $=$ $v_{0}$ $=$ (243 $\pm$ 1) $km~s^{-1}$, which leads to
$A$ $=$ $-$ $B$ $=$ 15.2 $km ~s^{-1} ~kpc^{-1}$ for $R$ $=$ 8 $kpc$
[the angular velocity is $\omega (R)$ $=$ $v_{0} / R$,
$A(R)$ $\equiv$ ($v(R)/R$ $-$ $dv(R)/dR$) / 2 $=$ $\omega (R)$ / 2,
in this special case when $v(R)$ $=$ $v_{0}$:
if $\omega (R)$ is a convex function,
then $A(R)$ is the convex function $\omega (R)$ / 2].

\section{Models of Galaxy}
Galaxy is standardly considered to consist of three components, galactic bulge,
galactic disk and galactic halo (see, e.g., Sec. 6 in Maoz 2007).
If one does not take into account
spiral structure of the Galaxy, then simple models can be created.

The model of Dauphole et al. (1996) considers
spherical symmetry for galactic bulge and halo and cylindrical symmetry
for galactic disk. Disadvantage of the model is that it does not
yield values of Oort constants corresponding to Eqs. (4).
Moreover, the model cannot be used as a realistic model
of Galaxy for galactocentric distances larger than about 40 kpc, since
the model produces a decreasing rotation curve for these distances.

We can create a new simple model, better consistent with
Eqs. (4) and with flat rotation curve of the Galaxy. The new model
considers galactic bulge of Dauphole et al. (1996) and its
gravitational potential is
\begin{eqnarray}\label{5}
\Phi_{bulge} ( r ) &=& -~ \frac{G~M_{b}}{\sqrt{r^{2} + b_{b}^{2}}} ~,
\nonumber \\
M_{b} &=& 1.3955 \times 10^{10} ~M_{\odot} ~,
\nonumber \\
b_{b} &=& 0.35 ~kpc ~,
\end{eqnarray}
where $G$ is the gravitational constant.

Galactic disk of our new model is represented by mass density function.
We can take relevant information from, e.g., Maoz (2007, Sec. 6.1.1.1):
``The Galactic disk has a mass distribution that falls exponentially
with both distance $R$ from the center and height $z$ above or below
the plane of the disk:
\begin{eqnarray}\label{6}
\varrho_{disk} ( R, z ) &=& \varrho_{0}~
     \left [ \exp \left ( - ~ \frac{R}{R_{d}} \right )	\right ]
     ~ \left [ \exp \left ( - ~ \frac{| z |}{h_{d}} \right )  \right ] ~.
\end{eqnarray}
The scale length of the disk, $R_{d}$ $=$ (3.5 $\pm$ 0.5) $kpc$, and hence
at $R$ $=$ 8 $kpc$ the Sun is in the outer regions of the Galaxy.
The characteristic scale height is $h_{d}$ $=$ 330 $pc$ for the lower-mass
(older) stars in the disk and $h_{d}$ $=$ 160 $pc$ for the gas-and-dust disk.
The Sun is located at $z$ $=$ 30 $pc$ above the midplane of the disk.
The mass of the disk within one scale radius is
$M_{disk} \approx 10^{10} M_{\odot}$, most of it is in stars (and about
10 $\%$ in gas).'' Eq. (3) is consistent with the statement presented by
Bertin (2000, p. 33): ``The luminosity profiles of galaxy disks are
approximately exponential, being reasonably well fitted by the law
$I(R)$ $=$ $I_{0}$ $\exp (- R / h)$, which has two scale parameters,
the central brightness $I_{0}$ and the exponential length $h$.''.

Finally, the simple model of galactic halo is given by a flat rotation curve.
The rotation curve is given by circular speed $v_{halo} (r)$ for spherical halo:
\begin{eqnarray}\label{7}
[ v_{halo} ( r ) ]^{2} &=& v_{H} ^{2} ~ \left \{
1 ~-~ \alpha ~\frac{a_{H}}{r} ~arctg \left ( \frac{r}{a_{H}} \right )
~-~ \left ( 1 - \alpha \right ) ~ \exp \left [ -~ \left (
    \frac{r}{b_{H}} \right )^{2} \right ] \right \} ~,
\nonumber \\
v_{H} &=& 220 ~km ~s^{-1} ~,
\nonumber \\
\alpha &=& 0.174 ~,
\nonumber \\
a_{H} &=& 0.04383 ~kpc ~,
\nonumber \\
b_{H} &=& 37.3760 ~kpc ~,
\end{eqnarray}
(compare with Sparke and Gallagher 2007, p. 95).
The numerical values of $\alpha$, $a_{H}$ and $b_{H}$ are calculated on the
basis of Eqs. (4)-(7) for the given value of $v_{H}$ (although a
different value can be used), $[ v_{halo} ( R ) ]^{2}$ $=$
$[ v_{Galaxy} ( R ) ]^{2}$ $-$ $[ v_{bulge} ( R ) ]^{2}$ $-$
$[ v_{disk} ( R ) ]^{2}$, $v_{Galaxy} ( R )$ $=$ $( A - B ) R$,
$dv_{Galaxy} ( R ) / d R$ $=$ $-$ $( A + B )$;
moreover, the values yield minimum
potential energy of the halo and also minimum radius of the Galaxy
for a given mass of the Galaxy.

\section{Motion in Galaxy -- general description}
Let us consider an approximation when global galactic gravitational
field is described by cylindrically symmetric potential $\Phi ( R, z )$,
$R$ being distance from the axis of rotation and $z$ the coordinate
of a body above/below the galactic equatorial plane ($z$ $=$ 0 corresponds
to the galactic equatorial plane; right-handed system $x-y-z$ has its origin
at the center of the Galaxy, $z$ is positively oriented toward the north
pole of the Galaxy; $R$ $=$ $\sqrt{x^{2} + y^{2}}$).
The galactic gravitational potential is generated by mass distribution
within the Galaxy. We can write
\begin{eqnarray}\label{8}
\Phi ( R, z ) &=& \Phi_{bulge} ( R, z ) ~+~
		  \Phi_{halo} ( R, z ) ~+~
		  \Phi_{disk} ( R, z ) ~.
\end{eqnarray}
Acceleration of the body is given by the following equations
in cartesian coordinates:
\begin{eqnarray}\label{9}
\frac{d^{2} X}{dt^{2}} &=& - ~ \frac{\partial \Phi ( R, z )}{\partial X} =
 - ~ \frac{\partial \Phi ( R, z )}{\partial R} ~
 \frac{X}{R} ~,
\nonumber \\
\frac{d^{2} Y}{dt^{2}} &=& - ~ \frac{\partial \Phi ( R, z )}{\partial Y} =
 - ~ \frac{\partial \Phi ( R, z )}{\partial R} ~
 \frac{Y}{R} ~,
\nonumber \\
\frac{d^{2} Z}{dt^{2}} &=& - ~ \left [
\frac{\partial \Phi ( R, z )}{\partial z} \right ]_{z=Z} ~,
\nonumber \\
R &=& \sqrt{X^{2} + Y^{2}} ~.
\end{eqnarray}
For the case represented by Eqs. (5)-(7) we have
\begin{eqnarray}\label{10}
\Phi_{bulge} ( R, z ) &=& \Phi_{bulge} ( r ) ~,
\nonumber \\
r &=& \sqrt{R^{2} + z^{2}} ~,
\end{eqnarray}
$\Phi_{disk} ( R, z )$ is found from the solution of Poisson's
equation
\begin{equation}\label{11}
\bigtriangleup ~\Phi_{disk} ( R, z ) = 4 ~\pi ~G~ \varrho_{disk} ( R, z ) ~,
\end{equation}
and, finally,
\begin{eqnarray}\label{12}
\Phi_{halo} ( R, z ) &=& \Phi_{halo} ( r ) ~,
\nonumber \\
r &=& \sqrt{R^{2} + z^{2}} ~,
\nonumber \\
\frac{d \Phi_{halo} \left ( r \right )}{dr} &=&
\frac{\left [ v_{halo} \left ( r \right ) \right ]^{2}}{r} ~.
\end{eqnarray}

\section{Motion near the galactic equator}
On the basis of Eqs. (5)-(7) and (10)-(12) we can write
\begin{eqnarray}\label{13}
\left [ \frac{\partial \Phi_{i}}{\partial R} \left ( R, z \right )
\right ]_{R_{0}}
&=&  \frac{\left [ v_{i} \left ( R_{0} \right ) \right ]^{2}}{R_{0}} ~
\left ( 1 ~+~ \alpha_{i} ~z^{2}  ~+~ \beta_{i} ~z^{4}  \right ) ~,
\nonumber \\
i &=& bulge, ~disk, ~halo ~,
\end{eqnarray}
and,
\begin{eqnarray}\label{14}
\left [ \frac{\partial \Phi}{\partial R} \left ( R, z \right ) \right ]_{R_{0}}
&=&  \sum_{i}
\left [ \frac{\partial \Phi_{i}}{\partial R} \left ( R, z \right )
\right ]_{R_{0}}
\nonumber \\
&=& \frac{v_{G}^{2}}{R_{0}} ~ \left \{ 1 ~+~
\frac{\sum_{i} v_{i}^{2}~\alpha_{i}}{v_{G}^{2}} ~z^{2} ~+~
\frac{\sum_{i} v_{i}^{2}~\beta_{i}}{v_{G}^{2}} ~z^{4} \right \} ~,
\nonumber \\
v_{G}^{2} &=& \sum_{i} v_{i}^{2} ~,
\nonumber \\
v_{G} &\equiv& v_{0} \equiv v ( R_{0} ) = ( A ~-~ B ) ~R_{0} ~.
\end{eqnarray}
Using numerical values for the model, we can write
\begin{eqnarray}\label{15}
\left [ \frac{\partial \Phi}{\partial R} \left ( R, z \right ) \right ]_{R_{0}}
&=& ( A ~-~ B )^{2} ~R_{0} ~ \left ( 1 ~-~ \Gamma_{1} ~z^{2} ~+~
\frac{1}{2} ~\Gamma_{2} ~z^{4} \right ) ~,
\nonumber \\
\Gamma_{1} &=& 0.124 ~kpc^{-2} ~,
\nonumber \\
\Gamma_{2} &=& 1.586 ~kpc^{-4} ~,
\nonumber \\
R_{0} &=& 8.0 ~kpc ~.
\end{eqnarray}
We have added also the value of $R_{0}$, the galactocentric distance of the Sun.
The values of $A$ and $B$ are given in Eqs. (4).

The rotation curve is given by the circular velocity
$v(R) = \sqrt{R ~\partial \Phi (R, 0) / \partial R}$. Eqs. (9) can be written,
then: $d^{2} X / dt^{2} = - ( v^{2} / R^{2} ) X$ and
$d^{2} Y / dt^{2} = - ( v^{2} / R^{2} ) Y$ for x- and y-components
in the plane $z =$ 0.
Using two close points, ($X_{0}$, $Y_{0}$, $Z_{0}$) and ($X$, $Y$, $Z$),
$X = X_{0} + \xi$, $Y = Y_{0} + \eta$, $Z = Z_{0} + \zeta$, one can write
$( X_{0} \xi + Y_{0} \eta + Z_{0} \zeta ) / \sqrt{R_{0}^{2} + Z_{0}^{2}}$
$\approx$ $( X_{0} \xi + Y_{0} \eta + Z_{0} \zeta ) / R_{0}$
for the difference between magnitudes of their position vectors,
if higher orders in $Z_{0} / R_{0}$,
$\xi$, $\eta$ and $\zeta$ are neglected. For the galactic plane
$[ v ( R ) ] ^{2} = v_{0}^{2} \{ 1 + 2 ( v_{0} ' / v_{0} )
( X_{0} \xi + Y_{0} \eta ) / R_{0} \}$,
where the prime denotes differentiation with respect to $R$
($v_{0} \equiv v(R_{0})$, $v_{0} ' \equiv [ dv(R) / dR]_{R_{0}}$)
and, again, higher orders in $\xi$ and $\eta$ are neglected.
We are dealing only with $| z |$ $\ll$ 1 $kpc$; $Z_{0}$ $=$ 0.03 $kpc$,
at present (Maoz 2007), $R_{0}$ $=$ 8 $kpc$.

The total action of all galactic components can be summarized
(see Eqs. 9):
\begin{eqnarray}\label{16}
\frac{d^{2} X}{dt^{2}} &=& - ~ \frac{v_{0}^{2}}{R_{0}^{2}}  \left \{
X_{0} + \xi  + 2 \left ( R_{0} \frac{v'_{0}}{v_{0}} - 1 \right )
\left [ \left ( \frac{X_{0}}{R_{0}} \right )^{2} \xi
+ \frac{X_{0} Y_{0}}{R^{2}_{0}} ~ \eta \right ] \right .
\nonumber \\
& & \left . -~X_{0}~ \left [ \Gamma_{1}
\left ( Z_{0}^{2} ~+~ 2~ Z_{0}~ \zeta \right ) ~-~
\frac{1}{2} ~\Gamma_{2}~
\left ( Z_{0}^{4} ~+~ 4 ~Z_{0}^{3}~ \zeta \right ) \right ]
 \right \} ~,
\nonumber \\
\frac{d^{2} Y}{dt^{2}} &=& - ~ \frac{v_{0}^{2}}{R_{0}^{2}}  \left \{
Y_{0} + \eta + 2 \left ( R_{0} \frac{v'_{0}}{v_{0}} - 1 \right )
\left [ \frac{X_{0} Y_{0}}{R^{2}_{0}} ~ \xi
+ \left ( \frac{Y_{0}}{R_{0}} \right )^{2} \eta \right ] \right .
\nonumber \\
& & \left . -~Y_{0}~ \left [ \Gamma_{1}
\left ( Z_{0}^{2} ~+~ 2~ Z_{0}~ \zeta \right ) ~-~
\frac{1}{2} ~\Gamma_{2}~
\left ( Z_{0}^{4} ~+~ 4 ~Z_{0}^{3}~ \zeta \right ) \right ]
 \right \} ~,
\end{eqnarray}
if higher orders in $\xi$, $\eta$, $\zeta$ are neglected.

\section{Relative acceleration}
The relative acceleration of the body with respect to the Sun, if 
($X_{0}$, $Y_{0}$, $Z_{0}$) represents position vector of the Sun, is:
$d^{2} \xi / dt^{2}$ $\equiv$ $-$ $G M_{\odot}	\xi / r^{3}$ $+$
$d^{2} X / dt^{2} - d^{2} X_{0} / dt^{2}$,
$d^{2} \eta / dt^{2}$ $\equiv$ $-$ $G M_{\odot}  \eta / r^{3}$ $+$
$d^{2} Y / dt^{2} - d^{2} Y_{0} / dt^{2}$, or
\begin{eqnarray}\label{17}
\frac{d^{2} \xi}{dt^{2}} &=& - ~ \frac{G M_{\odot}}{r^{3}} ~ \xi
\nonumber \\
& & - ~ \frac{v_{0}^{2}}{R_{0}^{2}}  \left \{
\left [ 1 - 2 \left ( \frac{X_{0}}{R_{0}} \right )^{2}
\left ( 1- R_{0} \frac{v'_{0}}{v_{0}} \right ) \right ] \xi ~-~ 2 ~
\frac{X_{0} Y_{0}}{R^{2}_{0}} \left ( 1- R_{0} \frac{v'_{0}}{v_{0}} \right )
~ \eta	\right .
\nonumber \\
& & \left . -~2~X_{0}~Z_{0}~ \left ( \Gamma_{1} ~-~ \Gamma_{2}~
Z_{0}^{2}~ \right ) ~ \zeta  \right \} ~,
\nonumber \\
\frac{d^{2} \eta}{dt^{2}} &=& - ~ \frac{G M_{\odot}}{r^{3}} ~ \eta
\nonumber \\
& & - ~ \frac{v_{0}^{2}}{R_{0}^{2}}  \left \{
\left [ 1 - 2 \left ( \frac{Y_{0}}{R_{0}} \right )^{2}
\left ( 1- R_{0} \frac{v'_{0}}{v_{0}} \right ) \right ] \eta ~-~ 2 ~
\frac{X_{0} Y_{0}}{R^{2}_{0}} \left ( 1- R_{0} \frac{v'_{0}}{v_{0}} \right )
~ \xi  \right .
\nonumber \\
& & \left . -~2~Y_{0}~Z_{0}~ \left ( \Gamma_{1} ~-~ \Gamma_{2}~
Z_{0}^{2}~ \right ) ~ \zeta  \right \} ~,
\nonumber \\
r &=& \sqrt{\xi ^{2} ~+~ \eta ^{2} ~+~ \zeta ^{2}} ~.
\end{eqnarray}
Using Oort constants $A$ and $B$, fulfilling the relations
$A$ $-$ $B$ $=$ $\omega_{0}$ $\equiv$ $v_{0} / R_{0}$,
$A$ $+$ $B$ $=$ $- v_{0}'$, Eq. (17) yields
\begin{eqnarray}\label{18}
\frac{d^{2} \xi}{dt^{2}} &=& - ~ \frac{G M_{\odot}}{r^{3}} ~ \xi
~+~ ( A - B ) \left [ A + B + 2 A \cos \left ( - 2 ~ \omega_{0} t \right )
 \right ] ~ \xi
\nonumber \\
& & +~ 2 A ( A - B ) \sin \left ( - 2 ~ \omega_{0} t \right ) ~\eta
\nonumber \\
& & +~ 2~(A - B)^{2} \left ( \Gamma_{1} ~-~ \Gamma_{2}~ Z_{0}^{2}~ \right )
~R_{0}~Z_{0}~ \cos \left ( - ~ \omega_{0} t \right ) ~ \zeta  ~,
\nonumber \\
\frac{d^{2} \eta}{dt^{2}} &=& - ~ \frac{G M_{\odot}}{r^{3}} ~ \eta
~+~  2 A ( A - B ) \sin \left ( - 2 ~ \omega_{0} t \right ) ~ \xi
\nonumber \\
& & +~ ( A - B ) \left [ A + B - 2 A \cos \left ( - 2 ~ \omega_{0} t \right )
 \right ] ~ \eta
\nonumber \\
& & +~ 2~(A - B)^{2} \left ( \Gamma_{1} ~-~ \Gamma_{2}~ Z_{0}^{2}~ \right )
~R_{0}~Z_{0}~ \sin \left ( - ~ \omega_{0} t \right ) ~ \zeta  ~,
\nonumber \\
r &=& \sqrt{\xi ^{2} ~+~ \eta ^{2} ~+~ \zeta ^{2}} ~,
\end{eqnarray}
where the sign minus at angular velocity ($-$ $\omega_{0}$) denotes
negative orientation of the galactic rotation
(clockwise orientation/direction of the solar motion with respect to the
center of the Galaxy), and, also the relations
$X_{0}$ $=$ $R_{0}$ $\cos ( - ~ \omega_{0} t )$ and
$Y_{0}$ $=$ $R_{0}$ $\sin ( - ~ \omega_{0} t )$ are used.

\subsection{$z-$component of the acceleration}
Eqs. (18) represent $x-$ and $y-$ components of acceleration.
In order to be the system of differential equations complete, we need
also the $z-$component of the acceleration. It can be obtained from the
Poisson's equation
\begin{eqnarray}\label{19}
\frac{1}{R} \frac{\partial}{\partial R} \left ( R ~
\frac{\partial \Phi}{\partial R} \right ) ~+~
\frac{\partial^{2} \Phi}{\partial z^{2}}
&=& 4 ~\pi ~G ~\varrho ( R, z ) ~.
\end{eqnarray}
Taking into account that $R \partial \Phi / \partial R$ $=$ $[v(R)]^{2}$ and
$[v(R)/R]_{R_{0}}$ $=$ $\omega_{0}$ $=$ $A$ $-$ $B$,
$[dv(R)/dR]_{R_{0}}$ $=$ $-$ ($A$ $+$ $B$), where $A$ and $B$ are
the Oort constants, and, the $z-$component of the acceleration is
$a_{z}$ $=$ $-$ $\partial \Phi / \partial z$, Eq. (19) yields
\begin{eqnarray}\label{20}
\frac{\partial a_{z}}{\partial z} &=& -~ 4 ~\pi ~G ~\varrho ( R_{0}, z ) ~-~
2~ ( A^{2} ~-~ B^{2} )
\nonumber \\
& & -~ 4 ~\pi ~G ~ \left [ \frac{\partial \varrho \left ( R, z \right )}{
    \partial R} \right ]_{R_{0}}
    \left \{ \frac{X_{0}}{R_{0}} ~ \left ( x - X_{0} \right ) ~+~
    \frac{Y_{0}}{R_{0}} ~ \left ( y - Y_{0} \right ) \right \} ~.
\end{eqnarray}
The first term in Eq. (14) is dominant for the region around the Sun.
Taking $\varrho (R_{0}, z)$ $=$ $\varrho(R_{0}, z_{0})$
in the right-hand side of Eq. (20), one finally obtains
\begin{eqnarray}\label{21}
a_{z} &=& -~ \left [ 4 ~\pi ~G ~\varrho ( R_{0}, z_{0} ) ~+~
2 \left ( A^{2} ~-~ B^{2} \right ) \right ] ~z
\nonumber \\
& & -~ 4 ~\pi ~G ~ \left [ \frac{\partial \varrho \left ( R, z_{0} \right )}{
    \partial R} \right ]_{R_{0}}
    \left \{ \frac{X_{0}}{R_{0}} ~ \left ( x - X_{0} \right ) ~+~
    \frac{Y_{0}}{R_{0}} ~ \left ( y - Y_{0} \right ) \right \} ~z ~.
\end{eqnarray}

We can take $\varrho ( R_{0}, z_{0} )$ as the sum of densities coming from
the disk, halo and bulge. The density coming from the bulge is of negligible
importance (see, e.g., the bulge model of Dauphole et al. 1996).
The contribution of the disk from the Maoz model
yields $\varrho_{disk}$ $=$ (0.126 $\pm$ 0.005)
$M_{\odot}$ $pc^{-3}$ for $z_{0}$ $=$ 0.
Considering spherically symmetric halo, its contribution to mass density is
$\varrho_{halo} ( r )$ $=$ ($4 \pi G$)$^{-1}$ $r^{-2}$
$d (r~ v_{halo}^{2}) / dr$, where $v_{halo}$ is circular velocity
of an object due to the gravity of the halo. One can use Eq. (7).
We show another possibility, now. We have
[$v_{Galaxy} (R)$]$^{2}$ $=$ [$v_{disk} (R)$]$^{2}$ $+$ [$v_{halo} (R)$]$^{2}$
$+$ [$v_{bulge} (R)$]$^{2}$, $v_{Galaxy} (R)$ $=$ [$A(R)$ $-$ $B(R)$] $R$,
$v_{disk} (R)$ $=$ [$A_{disk}(R)$ $-$ $B_{disk}(R)$] $R$.
Now, we have $\varrho_{halo} ( R_{0} )$ $=$ ($4 \pi G$)$^{-1}$ $R_{0}^{-1}$
$\{ R_{0}^{-1} v_{halo}^{2} - 2 ( A^{2} - B^{2} ) R_{0} +
2 ( A_{disk}^{2} - B_{disk}^{2} ) R_{0} - 2 v_{bulge}
[ d v_{bulge} (r) / dr ]_{R_{0}} \}$, if the relations
$[ d v_{Galaxy} (R) / dR ]_{R_{0}}$ $=$ $-$ ($A + B$),
$[ d v_{disk} (R) / dR ]_{R_{0}}$ $=$ $-$ ($A_{disk} + B_{disk}$) are used.
We have $\varrho_{halo}$ $\equiv$ $\varrho_{halo} ( R_{0} )$ $=$ 3.94 $\times$
10$^{-3}$ $M_{\odot}~pc^{-3}$. (As a comparison we can mention that the
halo mass density according to the halo model of Dauphole et al. (1996)
is 2.35-times greater than our result:
$\varrho_{halo} ( R_{0} )$ (Dauphole et al.) $=$ 9.26 $\times$ 10$^{-3}$
$M_{\odot}~pc^{-3}$.) We can summarize
\begin{eqnarray}\label{22}
[v_{Galaxy} (R)]^{2} &=& [v_{disk} (R)]^{2} + [v_{halo} (R)]^{2}
+ [v_{bulge} (R)]^{2} ~,
\nonumber \\
v_{Galaxy} (R) &=& [A(R) - B(R)] R ~,
\nonumber \\
v_{disk} (R) &=& [A_{disk}(R) - B_{disk}(R)] R ~,
\nonumber \\
v_{bulge} (r) &=& \sqrt{r ~\frac{\partial \Phi_{bulge} \left ( r \right )}{
		  \partial r}} ~,
\nonumber \\
\Phi_{bulge} (r) &\equiv& -~ \frac{G M_{b}}{\sqrt{r^{2} + b_{b}^{2}}} ~,
\nonumber \\
\varrho_{halo} ( R ) &=& (4 \pi G)^{-1} [ X(Galaxy) ~-~ X(disk) ~-~ X(bulge) ]	~,
\nonumber \\
X(Galaxy) &\equiv& -~ [A(R) - B(R)] \times [A(R) + 3 B(R)]
\nonumber \\
X(disk) &\equiv& -~ [A_{disk}(R) - B_{disk}(R)] \times [A_{disk}(R) + 3 B_{disk}(R)]
\nonumber \\
X(bulge) &\equiv& \frac{v_{bulge} \left ( R \right )}{R} ~  \left (
    \frac{v_{bulge} \left ( R \right )}{R} ~+~ 2 ~
    \frac{d v_{bulge} \left ( R \right )}{dR}  \right ) ~,
\nonumber \\
\sigma_{\varrho_{halo}} &=& \frac{2}{4 \pi G} ~ \times ~ \sqrt{Y} ~,
\nonumber \\
Y &\equiv& (A + B)^{2} \sigma_{A}^{2} ~+~ (A - 3 B)^{2} \sigma_{B}^{2} ~+~
\nonumber \\
& &	   (A_{disk} + B_{disk})^{2} \sigma_{A_{disk}}^{2} ~+~
	   (A_{disk} - 3 B_{disk})^{2} \sigma_{B_{disk}}^{2} ~,
\nonumber \\
M_{b} &\equiv& 1.3955 \times 10^{10} ~M_{\odot} ~,
\nonumber \\
b_{b} &\equiv& 0.35 ~kpc ~,
\nonumber \\
A_{disk} &\equiv& A_{disk} (R_{0}) = (10.81 \pm 1.70) ~km~ s^{-1}~ kpc^{-1} ~,
\nonumber \\
B_{disk} &\equiv& B_{disk} (R_{0}) = (-~9.97 \pm 0.34) ~km~ s^{-1}~ kpc^{-1} ~,
\nonumber \\
\varrho_{disk} &=& (0.126 \pm 0.005) ~ M_{\odot} ~ pc^{-3} ~,
\nonumber \\
\varrho_{halo} &=& (0.004 \pm 0.001) ~ M_{\odot} ~ pc^{-3} ~,
\nonumber \\
\varrho &=& \varrho_{disk} + \varrho_{halo}  ~,
\end{eqnarray}
if also the bulge model of Dauphole et al. (1996) is inserted.
The contribution of the bulge to the total mass density is
$\varrho_{bulge}$ / ($\varrho_{disk} + \varrho_{halo}$) $=$
1 $\times$ 10$^{-4}$.
The main part of the bulge is concentrated inside a sphere of small
radius and the value of $\varrho_{halo}$ is not very sensitive to the form
of the bulge potential, $v_{bulge}$ $\approx$ $\sqrt{GM_{b}/r}$.

The total value of the mass density, coming from the disk
and the halo, is $\varrho$ $=$ 0.130 $M_{\odot}$ $pc^{-3}$.
The value obtained from the model of Dauphole et al. (1996),
0.143 $M_{\odot}$ $pc^{-3}$, is in 10\% greater than our result.
The value of $\varrho$ seems to be in a good agreement with the values
(0.120 $\pm$ 0.008) $M_{\odot}$ $pc^{-3}$ and (0.138 $\pm$ 0.009) $M_{\odot}$
$pc^{-3}$ obtained from the results of Agekjan in 1962 and Agekjan and
Ogorodnikov in 1974 (see Kulikovskij 1985, pp. 158-159) under the assumption
that average mass of the stars in the solar neighborhood equals to
1 $M_{\odot}$ (it does not significantly differ from 1 $M_{\odot}$).

As for ($\partial \varrho / \partial R$)$_{R_{0}}$, the model of Maoz (2007)
yields [$\partial \varrho (R, z = 0) / \partial R$]$_{R_{0}~disk}$
$\equiv$ $\varrho'_{disk}$ $=$
$=$ ($-$0.0360 $\pm$ 0.0037) $M_{\odot} ~pc^{-3}~kpc^{-1}$ for $z_{0}$ $=$ 0.
Our model yields
[$\partial \varrho (r) / \partial r$]$_{R_{0}~halo}$ $\equiv$
$\varrho'_{halo}$ $=$ $-$ 0.62 $\times$ 10$^{-3}$
$M_{\odot} ~pc^{-3}~kpc^{-1}$.

Perhaps, the result represented by Eq. (21) can be a little improved using
the fact that
\begin{eqnarray}\label{23}
\varrho ( R_{0}, z ) &=& \varrho_{disk} ~ ( 1 ~-~ u ~| z | ) ~+~ \varrho_{halo} ~,
\nonumber \\
\left [ \frac{\partial \varrho \left ( R, z \right )}{
    \partial R} \right ]_{R_{0}} &=& \varrho'_{disk}  ~ ( 1 ~-~ u ~| z | ) ~+~
    \varrho'_{halo} ~,
\nonumber \\
\varrho_{disk} &=& (0.126 \pm 0.005) ~M_{\odot} ~pc^{-3} ~,
\nonumber \\
\varrho_{halo} &=& (0.004 \pm 0.001) ~M_{\odot} ~pc^{-3} ~,
\nonumber \\
\varrho_{disk}' &=& (-~ 0.0360 \pm 0.0037) ~M_{\odot} ~pc^{-3} ~kpc^{-1} ~,
\nonumber \\
\varrho_{halo}' &=& -~ 0.0006 ~M_{\odot} ~pc^{-3} ~kpc^{-1} ~,
\nonumber \\
u &=& 3.3 ~kpc^{-1} ~,
\end{eqnarray}
where the value of $u$ follows from the model of the disk.

Eqs. (20) and (23) are applied to both bodies, the one with the coordinates
($X_{0}$, $Y_{0}$, $Z_{0}$) and the second with the coordinates ($X$, $Y$, $Z$).
In application to the Solar System and the Oort cloud,
we have the Sun and the comet. We have
\begin{eqnarray}\label{24}
\frac{d^{2} Z_{0}}{dt^{2}} &=& -~ \left \{ 4 ~\pi ~G ~
\left [ \varrho_{disk} ~
\left ( 1 ~-~ \frac{1}{2} ~u ~ | Z_{0} | \right ) ~+~ \varrho_{halo} \right ] ~+~
2 \left ( A^{2} ~-~ B^{2} \right ) \right \} ~Z_{0} ~,
\nonumber \\
\frac{d^{2} Z}{dt^{2}} &=& -~ \left \{ 4 ~\pi ~G ~
\left [ \varrho_{disk} ~
\left ( 1 ~-~ \frac{1}{2} ~u ~ | Z_{0} | \right ) ~+~ \varrho_{halo} \right ] ~+~
2 \left ( A^{2} ~-~ B^{2} \right ) \right \} ~Z
\nonumber \\
& & -~ 4 ~\pi ~G ~
\left \{ \frac{X_{0}}{R_{0}} ~ \left ( X - X_{0} \right ) ~+~
	 \frac{Y_{0}}{R_{0}} ~ \left ( Y - Y_{0} \right ) \right \} \times
\nonumber \\
& & \times \left \{ \varrho'_{disk} ~
\left ( 1 ~-~ \frac{1}{2} ~ u~ | Z | \right ) ~+~ \varrho'_{halo} \right \} ~Z ~.
\end{eqnarray}
Initial conditions for the case of the Sun are:
$Z_{0}$ $=$ 30 $pc$, $dZ_{0}/dt$ $=$ $+$ 6 $km ~s^{-1}$
(information on the basic solar motion is used, see, e.g.,
Mihalas and McRae Routly 1968, p. 101).
Using $X$ $=$ $X_{0}$ $+$ $\xi$, $Y$ $=$ $Y_{0}$ $+$ $\eta$ and
$Z$ $=$ $Z_{0}$ $+$ $\zeta$, Eq. (24) yields
\begin{eqnarray}\label{25}
\frac{d^{2} \zeta}{dt^{2}} &=& - ~ \frac{G M_{\odot}}{r^{3}} ~ \zeta
\nonumber \\
& & -~ 4 ~\pi ~G \left \{ \varrho ~ \zeta ~-~ \varrho_{disk}
\frac{1}{2} ~u ~
\left [ | Z_{0} ~+~ \zeta | ~ \left ( Z_{0} ~+~ \zeta \right )
~-~ | Z_{0} | ~ Z_{0} \right ] \right \}
\nonumber \\
& & -~ 2 \left ( A^{2} ~-~ B^{2} \right ) ~\zeta
\nonumber \\
& & -~ 4 ~\pi ~G \left ( ~\varrho' ~-~ \varrho'_{disk} ~ \frac{1}{2} ~u ~
    | Z_{0} + \zeta | \right ) \left ( Z_{0} + \zeta \right )
    \left ( \frac{X_{0}}{R_{0}} ~ \xi ~+~
	 \frac{Y_{0}}{R_{0}} ~ \eta \right ) ~,
\end{eqnarray}
if also acceleration from the body with the coordinates
($X_{0}$, $Y_{0}$, $Z_{0}$) (Sun) is taken into account. The last term of
Eq. (25) assures that motion in a plane parallel to the plane of the galactic
equator does not exist: the inclination of the comet relative to the galactic
equatorial plane does not fulfill the relation $i$ $\equiv$ 0, since
$Z_{0}$ $\equiv$ 0 does not hold. The statement that
``plane-parallel motions do not exist'' is equivalent to the fact that
initial conditions $\zeta$ $=$ 0 and $d\zeta / dt$ $=$ 0 do not admit $\zeta$
$=$ constant $=$ 0 during the motion. Thus, our approach differs from the
approach of Bottlinger in 1924-1925 reducing to Oort equations for the galactic
equatorial plane (see Kulikovskij 1985, pp. 90-91).

\subsection{Summary}
Summarizing the most relevant equations and results presented above,
we obtain
\begin{eqnarray}\label{26}
\frac{d^{2} \xi}{dt^{2}} &=& - ~ \frac{G M_{\odot}}{r^{3}} ~ \xi
~+~ ( A - B ) ~\left [ A + B + 2 A \cos \left ( 2 ~ \omega_{0} t \right )
 \right ] ~ \xi
\nonumber \\
& & -~ 2~ A~ ( A - B )~ \sin \left ( 2 ~ \omega_{0} t \right ) ~\eta
\nonumber \\
& & +~ 2 ~( A - B )^{2} ~( \Gamma_{1} ~-~ \Gamma_{2} ~Z_{0}^{2} ) ~R_{0} ~Z_{0}
~\cos \left ( \omega_{0} t \right ) ~\zeta ~,
\nonumber \\
\frac{d^{2} \eta}{dt^{2}} &=& - ~ \frac{G M_{\odot}}{r^{3}} ~ \eta
~-~  2 ~A~ ( A - B )~ \sin \left ( 2 ~ \omega_{0} t \right ) ~ \xi
\nonumber \\
& & +~ ( A - B )~ \left [ A ~+~ B ~-~ 2 ~A ~\cos \left ( 2 ~ \omega_{0} t \right )
 \right ] ~ \eta
\nonumber \\
& & -~ 2 ~( A - B )^{2} ~( \Gamma_{1} ~-~ \Gamma_{2} ~Z_{0}^{2} ) ~R_{0} ~Z_{0}
~\sin \left ( \omega_{0} t \right ) ~\zeta ~,
\nonumber \\
\frac{d^{2} \zeta}{dt^{2}} &=& - ~ \frac{G M_{\odot}}{r^{3}} ~ \zeta
~-~ \left [ 4 ~\pi ~G ~\varrho ~+~
2 \left ( A^{2} ~-~ B^{2} \right ) \right ] ~\zeta
\nonumber \\
& & -~ 4 ~\pi ~G ~\varrho' ~
    Z_{0} \left [ \cos \left ( \omega_{0} t \right ) ~ \xi ~-~
    \sin \left ( \omega_{0} t \right ) ~ \eta \right ] ~,
\nonumber \\
\frac{d^{2} Z_{0}}{dt^{2}} &=& -~ \left [ 4 ~\pi ~G ~\varrho ~+~
2 \left ( A^{2} ~-~ B^{2} \right ) \right ] ~Z_{0} ~,
\nonumber \\
r &=& \sqrt{\xi ^{2} ~+~ \eta ^{2} ~+~ \zeta ^{2}} ~,
\nonumber \\
\omega_{0} &=& A ~-~ B ~,
\end{eqnarray}
if the terms containing the quantity $u$ are neglected and the relations
$X_{0}$ $=$ $R_{0}$ $\cos ( - ~ \omega_{0} t )$ and
$Y_{0}$ $=$ $R_{0}$ $\sin ( - ~ \omega_{0} t )$ are used.
The numerical values of the relevant quantities are:
\begin{eqnarray}\label{27}
A &=& ( 14.2 \pm 0.5 ) ~ km ~s^{-1} ~kpc^{-1} ~,
\nonumber \\
B &=&  ( -~ 12.4 \pm 0.5 ) ~ km ~s^{-1} ~kpc^{-1} ~,
\nonumber \\
\Gamma_{1} &=& 0.124 ~kpc^{-2} ~,
\nonumber \\
\Gamma_{2} &=& 1.586 ~kpc^{-4} ~,
\nonumber \\
\varrho &=& ( 0.130 \pm 0.005 ) ~M_{\odot}~ pc^{-3} ~,
\nonumber \\
\varrho' &=& ( -~ 0.037 \pm 0.004 ) ~M_{\odot} ~pc^{-3} ~kpc^{-1} ~.
\end{eqnarray}
We stress that $\varrho$ $\equiv$ $\varrho ( R_{0}, z = 0 )$ $=$
0.130 $M_{\odot}~ pc^{-3}$, $R_{0}$ $=$ 8.0 $kpc$. The mass density
in the solar neighborhood, 30 $pc$ above the galactic equatorial plane, is
$\varrho$ $\equiv$ $\varrho ( R_{0}, z = 0.03~kpc )$ $=$ $(0.117 \pm 0.005)$
$M_{\odot}~ pc^{-3}$, see also Eq. (23).

One must bear in mind that the values of $A$ and $B$
significantly differ from the IAU recommended values
$A$ $=$ 15 ~ $km ~s^{-1} ~kpc^{-1}$,
$B$ $=$ $-$ 10~ $km ~s^{-1} ~kpc^{-1}$.
The following results should be for certain:
\begin{eqnarray}\label{28}
A &\in& \langle 10.0, 15.0 \rangle ~ km ~s^{-1} ~kpc^{-1} ~,
\nonumber \\
B &\in&  \langle -~ 15.0, -~ 10.0 \rangle  ~ km ~s^{-1} ~kpc^{-1} ~,
\nonumber \\
\varrho &\in& ( 0.07, 0.15 )  ~M_{\odot}~ pc^{-3} ~,
\nonumber \\
\varrho ' &\in& ( - ~0.043, -~ 0.035 ) ~M_{\odot} ~pc^{-3} ~kpc^{-1} ~,
\end{eqnarray}
if also the result for the mass density of Cr\'{e}z\'{e} et al. (1998)
and the IAU recommended values are considered. Unfortunately, there
exist results which do not fit Eq. (28) (e.g., there are values of $A$ and $B$
which are out of the above presented intervals, see Clemens (1985)).

\section{Galactic tide for Dauphole et al. (1996) model of Galaxy}
Taking into account Dauphole et al. (1996) model of Galaxy,
the relevant equations for galactic tide are:
\begin{eqnarray}\label{29}
\frac{d^{2} \xi}{dt^{2}} &=& - ~ \frac{G M_{\odot}}{r^{3}} ~ \xi
~+~ ( A - B ) ~\left [ A + B + 2 A \cos \left ( 2 ~ \omega_{0} t \right )
 \right ] ~ \xi
\nonumber \\
& & -~ 2~ A~ ( A - B )~ \sin \left ( 2 ~ \omega_{0} t \right ) ~\eta
\nonumber \\
& & +~ ( A - B )^{2} ~( \Gamma_{1D} / \sqrt{b_{d}^{2} + Z_{0}^{2}}
 ~+~ \Gamma_{2D} ) ~R_{0} ~Z_{0}
~\cos \left ( \omega_{0} t \right ) ~\zeta ~,
\nonumber \\
\frac{d^{2} \eta}{dt^{2}} &=& - ~ \frac{G M_{\odot}}{r^{3}} ~ \eta
~-~  2 ~A~ ( A - B )~ \sin \left ( 2 ~ \omega_{0} t \right ) ~ \xi
\nonumber \\
& & +~ ( A - B )~ \left [ A ~+~ B ~-~ 2 ~A ~\cos \left ( 2 ~ \omega_{0} t \right )
 \right ] ~ \eta
\nonumber \\
& & -~ ( A - B )^{2} ~( \Gamma_{1D}  / \sqrt{b_{d}^{2} + Z_{0}^{2}}
~+~ \Gamma_{2D} ) ~R_{0} ~Z_{0}
~\sin \left ( \omega_{0} t \right ) ~\zeta ~,
\nonumber \\
\frac{d^{2} \zeta}{dt^{2}} &=& - ~ \frac{G M_{\odot}}{r^{3}} ~ \zeta
~-~ \left [ 4 ~\pi ~G ~\varrho ~+~
2 \left ( A^{2} ~-~ B^{2} \right ) \right ] ~\zeta
\nonumber \\
& & -~ 4 ~\pi ~G ~\varrho' ~
    Z_{0} \left [ \cos \left ( \omega_{0} t \right ) ~ \xi ~-~
    \sin \left ( \omega_{0} t \right ) ~ \eta \right ] ~,
\nonumber \\
\frac{d^{2} Z_{0}}{dt^{2}} &=& -~ \left [ 4 ~\pi ~G ~\varrho ~+~
2 \left ( A^{2} ~-~ B^{2} \right ) \right ] ~Z_{0} ~,
\nonumber \\
r &=& \sqrt{\xi ^{2} ~+~ \eta ^{2} ~+~ \zeta ^{2}} ~,
\nonumber \\
\omega_{0} &=& A ~-~ B ~,
\nonumber \\
A &=& 14.25 ~ km ~s^{-1} ~kpc^{-1} ~,
\nonumber \\
B &=&  -~ 13.89 ~ km ~s^{-1} ~kpc^{-1} ~,
\nonumber \\
\Gamma_{1D} &=& 0.084 ~kpc^{-1} ~,
\nonumber \\
\Gamma_{2D} &=& 0.008 ~kpc^{-2} ~,
\nonumber \\
\varrho &=& 0.143 ~M_{\odot}~ pc^{-3} ~,
\nonumber \\
\varrho' &=& -~ 0.0425 ~M_{\odot} ~pc^{-3} ~kpc^{-1} ~,
\nonumber \\
b_{d} &=& 0.25 ~kpc ~.
\end{eqnarray}

\section{Conclusion}
The paper treats the effect of the galactic tide on the relative motion
of two close bodies moving in the galactic disk. Equations considering
the effect of the galactic tide on motion of a body near the Sun are
given by Eqs. (26). Values of the physical quantities are presented in
Eqs. (23) and (27), or, the more rough values are given by Eqs. (28).
Taking into account Dauphole et al. (1996) model of Galaxy,
equations for galactic tide are also presented.

The effect of the normal $z-$component is conventional.
The $x-$ and $y-$ components of the acceleration comes not only
from the $x-$ and $y-$ components of the position of a body, but also from
its $z-$component of the position. Motion of the Sun normal to the galactic
equator plays also an important role.

The physical models presented here will be used in a forthcoming paper
for a more realistic study of orbital evolution of long-period comets.

\section*{Acknowledgement}
This work was supported by the Scientific Grant Agency VEGA, Slovak Republic,
grant No. 2/0016/09.

\end{document}